\documentclass[submission,copyright,creativecommons]{eptcs}
\providecommand{\event}{ICLP 2026} 

\usepackage{iftex}
\usepackage{amsmath}
\usepackage{amsfonts}
\usepackage{graphicx}
\usepackage{multirow}
\usepackage{listings}
\usepackage{algorithm}
\usepackage{algpseudocode}
\usepackage{tikz}
\usepackage{tikz-qtree}
\usetikzlibrary{automata, positioning, arrows}
\usetikzlibrary{positioning}
\usepackage{subfigure}
\usepackage{xspace}
\usepackage{comment}
\usepackage{caption}
\usepackage{siunitx}
\usepackage{url}

\newcommand{\METHOD}{STRADA\xspace}

\lstdefinelanguage{asp}{
    morekeywords={color, cl, seq, item, patpos, patlen, pat, occ, seqlen, support, card, sup, gr, csp, rocc, ins, occS, occG, occSG, trace, total, node, label, edge, order, op, event, var, binding, holds, last, sat, unsat, supp, supported, confident, node, edge, startn, endn, length, headway, flight, lane, start, end, speed, requested, eta, stpoint, new_flight, overtake, delay, sum_delay, way, reachable, before, dl, left_border, right_border, size},
    sensitive=true, 
    comment=[l]
    morestring=[b]',     
    keywordstyle=\color{blue}, 
    commentstyle=\color{green}, 
    stringstyle=\color{red}, 
    numbers=left,
    numberstyle=\tiny\color{gray},
    frame=tb
}

\ifpdf
  \usepackage{underscore}         
  \usepackage[T1]{fontenc}        
\else
  \usepackage{breakurl}           
\fi

\title{Declarative Problem Solving in UAM Strategic Deconfliction}
\author{Gioacchino Sterlicchio
\institute{DMMM, Polytechnic University of Bari\\Bari, Italy}
\email{g.sterlicchio@phd.poliba.it}
\and
Angelo Oddi 
\institute{ISTC-CNR\\
Rome, Italy}
\email{angelo.oddi@istc.cnr.it}
\and
Riccardo Rasconi
\institute{ISTC-CNR\\
Rome, Italy}
\email{riccardo.rasconi@istc.cnr.it}
\and
Francesca Alessandra Lisi
\institute{DIB and CILA, University of Bari Aldo Moro\\Bari, Italy}
\email{FrancescaAlessandra.Lisi@uniba.it}
}

\def\titlerunning{Declarative Problem Solving in UAM Strategic Deconfliction}
\def\authorrunning{G. Sterlicchio, A. Oddi, R. Rasconi \& F.A. Lisi}
\begin{document}
\maketitle

\begin{abstract}

The growing demand for Urban Air Mobility (UAM) introduces significant challenges in airspace management, particularly within densely populated metropolitan regions. 
As the number of aerial vehicles---such as drones, air taxis, and helicopters---continues to rise, so does the risk of mid-air collisions and conflicts with existing air traffic and obstacles. Ensuring safe and efficient UAM operations requires robust strategic deconfliction mechanisms.
We propose an Answer Set Programming (ASP) based approach for strategic deconfliction, focusing on time synchronization and route optimization for conflict-free flight plans. The solution is benchmarked against Constraint Programming (CP), emphasizing scalability and resource use. Results show that ASP offers faster execution and better scalability for small to medium cases, while CP maintains stable memory but degrades with complexity. 
\end{abstract}

\section{Introduction}\label{sect:sd-intro}

Urban Air Mobility (UAM) is emerging as a transformative solution to mitigate urban traffic congestion, reduce emissions, and enhance transportation efficiency \cite{Pak2024}. The UAM ecosystem includes various aerial vehicles based on electric Vertical Take-Off and Landing (eVTOL), such as drones, intended for operations in densely populated areas. Integrating these vehicles into existing airspace poses several issues related to safety, efficiency, and scalability. 
To address these issues, industry and regulators are revising UAM ConOps by designating dedicated air lanes for eVTOLs, reducing reliance on traditional Air Traffic Control \cite{fed20}. 
However, a key challenge remains the prediction, detection, and resolution of potential conflicts among air vehicles, other traffic, obstacles, and restricted zones \cite{xue2020urban}.
This paper extends a common framework to develop models and algorithms for UAM \textit{Strategic Deconfliction} (SD), the first of three layers in Air Traffic Management \cite{liu2023strategic}. SD addresses conflict detection and resolution prior to departure through resource allocation, airspace organization, demand-capacity balancing, and traffic synchronization. Subsequent layers include \textit{Tactical Deconfliction}, which ensures in-flight separation using real-time data, and \textit{Collision Avoidance}, which relies on onboard systems to prevent imminent collisions. Our focus is on traffic synchronization within SD, emphasizing proactive trajectory planning to reduce conflicts and optimize airspace utilization, supported by advanced algorithms and predictive analytics.

We tackle the SD problem using Answer Set Programming (ASP), a declarative paradigm well-suited for complex combinatorial optimization. ASP enables modelling intricate constraints in air traffic management and efficiently computing conflict-free trajectories under safety and operational requirements \cite{DBLP:journals/aim/Lifschitz16}. Our proposed framework, \METHOD (Scalable Trajectory Resolution Architecture for Deconfliction using ASP), represents the first logic-based approach in this domain, contrasting with existing numerical methods. Preliminary results obtained with a former version of \METHOD have been presented in \cite{DBLP:conf/cilc/SterlicchioORL25}. In this paper, 
a more extensive evaluation of \METHOD focuses on scalability while varying flight volumes, airspace sizes, and take-off intervals, and includes a comparative analysis with an encoding in Constraint Programming (CP). CP is a declarative method, widely used in scheduling and optimization, for solving complex combinatorial problems by defining variables, their possible values (domains), and the rules (constraints) they must follow, letting to find solutions that satisfy all conditions \cite{DBLP:reference/fai/2}. 
The contributions of the paper are threefold: (1) an ASP formalisation of the SD problem,
(2) the \METHOD framework which consists of the air network topology modelling, the fleet of drones to be organised as well as the SD encoding, 
and (3) the empirical evaluation of the efficiency of \METHOD through scalability tests aimed at assessing its strengths and weaknesses, and comparing them against CP.

The paper is structured as follows.
Section~\ref{sect:sd-rel-works} reviews recent literature on the SD problem in UAM as well as ASP applications that can be considered related to our work. 
Section~\ref{sect:preliminaries} introduces ASP and CP fundamentals.
Section~\ref{sect:problem statement} formalizes the SD problem with a practical example, 
while Section~\ref{sect:asp solution} details the \METHOD framework and its modelling rationale. 
Section~\ref{sect:sd-evaluation} presents the evaluation methodology focusing on execution time and memory usage. 
Finally, Section~\ref{sect:sd-conclusions} summarizes the findings and outlines future research directions.
\vspace{-5mm}
\section{Related Works}\label{sect:sd-rel-works}
Recent research on SD in UAM explores diverse approaches for conflict-free trajectory planning and airspace management.
In \cite{sacharny2019lane} 
an airspace structure inspired by roadway roundabouts, and a computationally tractable trajectory scheduling algorithm for UAS Service Suppliers are presented. 
Another work addresses the design of a parcel delivery system using drones including strategic planning made by mixed integer linear programming \cite{DBLP:journals/jirs/TorabbeigiLK20}.
Some authors proposed a mixed-integer Second-Order Cone program, a convex programming model that has second-order constraints \cite{tang2021automated}.  
Another work contributes on an efficient lane-based strategic deconfliction scheduling algorithm and a tactical deconfliction protocol to handle dynamic contingencies (e.g., failure to follow the nominal flight plan) \cite{DBLP:journals/tits/SacharnyHM22}.
In \cite{huang2022strategic}, the SD problem in UAM is addressed with multi-agent reinforcement learning.
A framework for generating routes leveraging a rapidly-exploring random tree based algorithm \cite{thompson2023framework} is also proposed. 
A reinforcement learning approach and mathematical programming for strategic conflict management \cite{DBLP:journals/tits/ChenEBW24} is also used.
Another work introduces a stacked hexagonal tessellation to model the airspace and an optimization-based procedure based on integer programming for SD \cite{liu2023strategic}.

ASP has not yet been used in SD. Therefore, we mention here scheduling works that use ASP in seemingly equivalent application domains. 
Examples include employee team generation in seaports \cite{DBLP:journals/tplp/RiccaGAMLIL12}, nurse scheduling and rescheduling with optimal shift assignments \cite{DBLP:conf/lpnmr/DodaroM17,DBLP:journals/ia/AlvianoDM18}, and chemotherapy appointment scheduling in oncology clinics \cite{DBLP:journals/tplp/DodaroGGMMP21}. ASP has also been used for complex train scheduling tasks through hybrid approaches incorporating difference constraints via \verb|clingo[DL]| \cite{DBLP:conf/lpnmr/AbelsJOSTW19,DBLP:journals/tplp/AbelsJOSTW21,DBLP:journals/tplp/JanhunenKOSWS17}. Further applications include warehouse delivery scheduling, ensuring collision-free robot movements and task completion \cite{DBLP:journals/algorithms/RajaratnamSWCLS23}.
\section{Preliminaries on ASP and CP}
\label{sect:preliminaries}
Answer Set Programming (ASP) \cite{DBLP:journals/aim/Lifschitz16} is a declarative problem solving paradigm, where a problem in encoded as a logic program such that its answer sets or stable models \cite{DBLP:conf/iclp/GelfondL88} correspond to the solutions of the problem and are computable using ASP solvers, e.g., \verb|clingo| \cite{gebser2014clingo}.
An ASP program is a finite set of rules of the form $a \leftarrow \;b_1,\dots,b_m,\mathit{not}\;b_{m+1},\dots,\mathit{not}\;b_n$ where $a$ and all $b_i$ and are first-order atoms of form $p(t_1,\dots,t_k)$ and all $t_i$ are terms, composed of function symbols and variables. The atom $a$ is often called head atom, while $b_1,\dots,b_m$ and $\mathit{not}\;b_{m+1},\dots,\mathit{not}\;b_n$ are also referred to as positive and negative body, respectively. An expression is said to be ground, if it contains no variables.
As usual, $\mathit{not}$ denotes default negation. A rule is called a fact if $m=n=0$ and an integrity constraint if there is no head. In what follows, we deal with normal logic programs only, for which we can have rules with only one atom in the head or without head.

Constraint Programming (CP) \cite{DBLP:reference/fai/2} is an alternative approach to programming which relies on a combination of techniques that deal with reasoning and computing . It has been successfully applied in a number of fields including molecular biology, electrical engineering, operations research and numerical analysis.
Formally, a problem is defined as a triple $(X,D,C)$, where: 
$X=\{x_1,\dots,x_n\}$ is the set of variables of the problem;
$D=\{D_1,\dots,D_n\}$ is the set of domains of the variables, i.e., for all $k \in [1;n]$ we have $x_k \in D_k$ and
$C=\{C_1,\dots,C_m\}$ is a set of constraints. 
A constraint $C_i=(X_i,R_i)$ is defined by a set $X_i={x_{i_1},\dots x_{i_k}}$ of variables and a relation $R_i \subseteq D_i \times \dots \times D_{i_k}$ that defines the set of values allowed simultaneously for the variables of $X_i$.
An evaluation is consistent if it does not violate any of the constraints. An evaluation is complete if it includes all variables. An evaluation is a solution if it is consistent and complete; such an evaluation is said to solve the problem.
\section{The Strategic Deconfliction Problem}\label{sect:problem statement}

\subsection{Lane-based airspace structure}\label{subsect:lane based}
\begin{figure}[ht]
    \centering
    \subfigure[]{
    \begin{tikzpicture}[node distance={15mm}, thick, main/.style = {draw, circle}] 
    \node[main] (1) [] {1}; 
    \node[main] (2) [right of=1] {2}; 
    \node[main] (3) [right of=2] {3}; 
    \node[main] (4) [right of=3] {4};  
    \draw[->] (1) -- (2); 
    \draw[->] (2) -- (3);
    \draw[->] (3) -- (4); 
    \end{tikzpicture}
    }
    \hspace{5mm}
    \subfigure[]{
    \begin{tikzpicture}[node distance={15mm}, thick, main/.style = {draw, circle}] 
    \node[main] (1) {1}; 
    \node[main] (2) [right of=1] {2}; 
    \node[main] (3) [right of=2] {3}; 
    \node[main] (4) [right of=3] {4};
    
    \draw[->] (1.20) -- (2.160);
    \draw[->] (2.200) -- (1.340);
    \draw[->] (2.20) -- (3.160);
    \draw[->] (3.200) -- (2.340);
    \draw[->] (3.20) -- (4.160);
    \draw[->] (4.200) -- (3.340);
    \end{tikzpicture}
    }
    \caption{A simple one-way (a) and two-way (b) three-lane layout}
    \label{fig:lane-layout}
\end{figure}
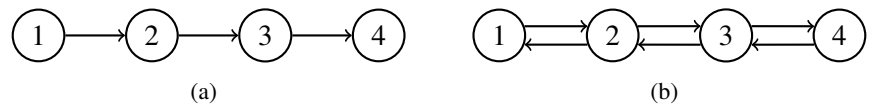

The proposed approach adopts a lane-based airspace structure \cite{DBLP:journals/tits/SacharnyHM22}, where lanes are modelled as directed graphs with vertices representing entry/exit points and edges representing one-way corridors. Two-way traffic between vertexes can be achieved by having pairs of air lanes next to each other at the same altitude or at different altitudes.
Compared to FAA-NASA strategic deconfliction (FNSD), lane-based methods reduce computational complexity but impose limitations such as increased travel distance and frequent turns \cite{sacharny2020faa}. Despite these drawbacks, lanes enable efficient real-time deconfliction and contingency management \cite{sacharny2020dddas}. Lanes form 3D corridors with design constraints like headway for safe separation and may include properties such as speed limits. 
The combination of headway requirements and corridor design supports a variety of vehicle trajectory constraints, while the directed graph structure imposed on the airspace provides agents with an organized environment for computation.
This structured approach contrasts with zone-based and cell-based deconfliction, offering a more organized and computationally tractable environment for trajectory planning.
Figure~\ref{fig:lane-layout} shows a lane structure as a graph $G=(V,E)$, where $V=\{1,2,3,4\}$ and $E=\{(1,2), (2,3), (3,4)\}$ for the one-way structure and $E=\{(1,2),(2,1),(2,3),(3,2),(3,4),(4,3)\}$ for the two-way alternative. In particular, vertex $1$ and $4$ are ground nodes, while vertex $2$ and $3$ are waypoints  at some altitude. The lanes allow only one direction of travel without overtaking.

\subsection{Problem introduction}\label{subsect:problem introduction}
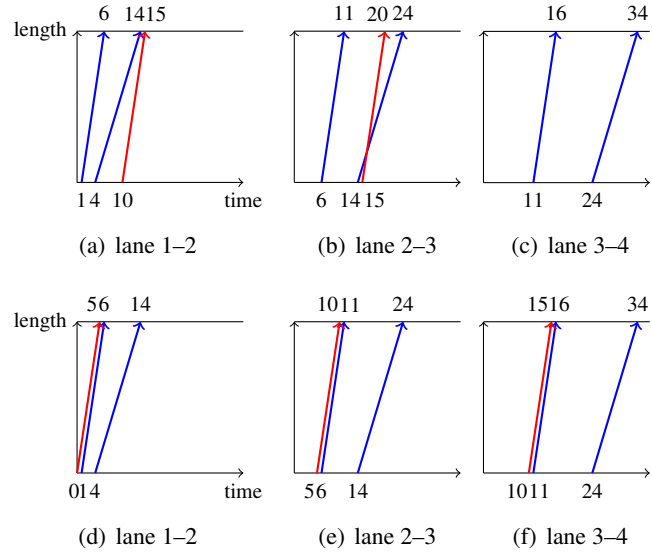
\begin{figure*}[ht]
    \centering
    \subfigure[lane 1--2]{
    \begin{tikzpicture}[scale=0.20]
    \draw[->] (0,0) -- (11,0) node[right, below] {\scriptsize time};
    \draw[->] (0,0) -- (0,10) node[left] {\scriptsize length};
    \draw[] (0,10) -- (11,10);

    \draw (0.3,0) node[right, below, black] {\scriptsize 1};
    \draw[blue, thick, ->] (0.3,0) -- (1.8,10) node[right, above, black] {\scriptsize 6};
    
    \draw (1.2,0) node[right, below, black] {\scriptsize 4};
    \draw[blue, thick, ->] (1.2,0) -- (4.2,10);
    \draw (3.8,10) node[right, above, black] {\scriptsize 14};
    
    \draw (3,0) node[right, below, black] {\scriptsize 10};
    \draw[red, thick, ->] (3,0) -- (4.5,10); 
    \draw(5.2,10) node[right, above, black] {\scriptsize 15};
    \end{tikzpicture}
    }
    \subfigure[lane 2--3]{
    \begin{tikzpicture}[scale=0.20]
    \draw[->] (0,0) -- (11,0);
    \draw[->] (0,0) -- (0,10);
    \draw[] (0,10) -- (11,10);

    \draw (1.8,0) node[right, below, black] {\scriptsize 6};
    \draw[blue, thick, ->] (1.8,0) -- (3.3,10) node[right, above, black] {\scriptsize 11};
    
    \draw (3.7,0) node[right, below, black] {\scriptsize 14};
    \draw[blue, thick, ->] (4.2,0) -- (7.2,10) node[right, above, black] {\scriptsize 24};
    
    \draw (5.3,0) node[right, below, black] {\scriptsize 15};
    \draw[red, thick, ->] (4.5,0) -- (6,10); 
    \draw(5.5,10) node[right, above, black] {\scriptsize 20};
    \end{tikzpicture}
    }
    \subfigure[lane 3--4]{
    \begin{tikzpicture}[scale=0.20]
    \draw[->] (0,0) -- (11,0);
    \draw[->] (0,0) -- (0,10);
    \draw[] (0,10) -- (11,10);

    \draw (3.3,0) node[right, below, black] {\scriptsize 11};
    \draw[blue, thick, ->] (3.3,0) -- (4.8,10) node[right, above, black] {\scriptsize 16};
    
    \draw (7.2,0) node[right, below, black] {\scriptsize 24};
    \draw[blue, thick, ->] (7.2,0) -- (10.2,10) node[right, above, black] {\scriptsize 34};
    \end{tikzpicture}
    }
    \\
    \subfigure[lane 1--2]{
    \begin{tikzpicture}[scale=0.20]
    \draw[->] (0,0) -- (11,0) node[right, below] {\scriptsize time};
    \draw[->] (0,0) -- (0,10) node[left] {\scriptsize length};
    \draw[] (0,10) -- (11,10);

    \draw (0.4,0) node[right, below, black] {\scriptsize 1};
    \draw[blue, thick, ->] (0.3,0) -- (1.8,10) node[right, above, black] {\scriptsize 6};
    
    \draw (1.2,0) node[right, below, black] {\scriptsize 4};
    \draw[blue, thick, ->] (1.2,0) -- (4.2,10) node[right, above, black] {\scriptsize 14};
    
    \draw (-0.2,0) node[right, below, black] {\scriptsize 0};
    \draw[red, thick, ->] (0,0) -- (1.5,10); 
    \draw(1,10) node[right, above, black] {\scriptsize 5};
    \end{tikzpicture}
    }
    \subfigure[lane 2--3]{
    \begin{tikzpicture}[scale=0.20]
    \draw[->] (0,0) -- (11,0);
    \draw[->] (0,0) -- (0,10);
    \draw[] (0,10) -- (11,10);

    \draw (1.8,0) node[right, below, black] {\scriptsize 6};
    \draw[blue, thick, ->] (1.8,0) -- (3.3,10);
    \draw (3.7,10) node[right, above, black] {\scriptsize 11};
    
    \draw (4.2,0) node[right, below, black] {\scriptsize 14};
    \draw[blue, thick, ->] (4.2,0) -- (7.2,10) node[right, above, black] {\scriptsize 24};
    
    \draw (1,0) node[right, below, black] {\scriptsize 5};
    \draw[red, thick, ->] (1.5,0) -- (3,10);
    \draw (2.2,10) node[right, above, black] {\scriptsize 10};
    \end{tikzpicture}
    }
    \subfigure[lane 3--4]{
    \begin{tikzpicture}[scale=0.20]
    \draw[->] (0,0) -- (11,0);
    \draw[->] (0,0) -- (0,10);
    \draw[] (0,10) -- (11,10);

    \draw (3.7,0) node[right, below, black] {\scriptsize 11};
    \draw[blue, thick, ->] (3.3,0) -- (4.8,10);
    \draw (5,10) node[right, above, black] {\scriptsize 16};
    
    \draw (7.2,0) node[right, below, black] {\scriptsize 24};
    \draw[blue, thick, ->] (7.2,0) -- (10.2,10) node[right, above, black]  {\scriptsize 34};
    \draw (2.2,0) node[right, below, black] {\scriptsize 10};
    \draw[red, thick, ->] (3,0) -- (4.5,10);
    \draw (3.6,10) node[right, above, black] {\scriptsize 15};
    \end{tikzpicture}
    }
    \caption{STDL representation of the flights. Conflict case (a--c), no conflict case (d--f)}
    \label{fig:stdl-diagram}
\end{figure*}

The SD problem is to produce a set of scheduled flight paths such that no two aircraft ever get closer than a specified safety distance or \emph{headway} $h$ either in time or space.
Consider Figure~\ref{fig:lane-layout},
a flight must schedule its entry-exit times through a sequence of lanes, where the exit time from the previous lane equals the entry time of the following lane.
In order to determine whether flights have a conflict, we use the Space-Time Lane Diagram (STLD) \cite{DBLP:journals/tits/SacharnyHM22} to represent the situation graphically, as shown in Figure~\ref{fig:stdl-diagram}. 
The horizontal axis represents time, while the vertical axis represents the distance along the lane.
A STDL is created for each lane. The two blue lines represent two scheduled flights $f_1$ and $f_2$ with start times of 1 and 4 in lane 1--2 with speeds 2 and 1, respectively.
The STDL shows their progress through the three lanes; it can be seen that there is always a time headway of at least 1 unit.
Suppose a new flight $f_3$ must be scheduled, with speed 2, and the requested launch interval is $[0,21]$. This means that the earliest launch time is $0$ while the latest one is $21$.
The goal is to establish departure times that do not conflict with those already present and find the trajectory for all lanes that does not conflict with all flights travelling in their respective lanes.
For example, consider $f_3$ starting at time 10 (red line); then it exits Lane 1-2 and enters Lane 2-3 at time 15; the figure shows that $f_3$ crosses the path of $f_2$ and therefore is disallowed.
On the other hand, if $f_3$ starts at time 0, then its headway is always equal to 1 time unit from $f_1$, and since both flights are characterized by the same speed, they never get any closer. Moreover, for Lane 2--3 and Lane 3--4 $f_3$'s headway is still 1 unit from $f_1$, so it is allowed.

\subsection{Problem formalization}\label{subsect:problam formalization}
We formalize the SD problem as a couple $(N,F)$ where $N$ is the airspace network and $F$ is the set of flights to be scheduled in $N$. 
The network $N$ is in its turn a tuple $(V,E,\mathit{Iv},\mathit{Ev},l,h)$, where: 
$(V,E)$ is a directed graph,
$\mathit{Iv} \subset V$ is the set of ground nodes where a flight starts its trip,
$\mathit{Ev} \subset V$ is the set of ground nodes where a flight ends its trip,
$l:E\rightarrow \mathbb{N}$ assigns the lane length and
$h \in \mathbb{N}$ is the headway distance (time or space) between flights.
The set $F$ is represented by tuples  $f_i = (S_i,L_i,\mathit{start_i},\mathit{end_i},e_i,l_i,s_i)$, where:
$(S_i,L_i)$ is an acyclic sub-graph of $(V,E)$ and represents the flight route,
$\mathit{start_i}~:F\rightarrow \mathit{Iv}$ gives the ground vertex where a flight starts its trip,
$\mathit{end_i}~:F\rightarrow \mathit{Ev}$ gives the ground vertex where a flight ends its trip,
$e_i:F\rightarrow \mathbb{N}$ gives the earliest time a flight can start its trip,
$l_i:F\rightarrow \mathbb{N}$ gives the latest time a flight can start its trip and
$s_i:F\rightarrow \mathbb{N}$ is the speed associated to a flight.
Following Figure~\ref{fig:lane-layout}~(a), $N$ is defined as: $V=\{1,2,3,4\}$, 
$E=\{(1,2),(2,3),(3,4)\}$, 
$\mathit{Iv}=\{1\}$, 
$\mathit{Ev}=\{4\}$, 
$l(1,2)=10$, $l(2,3)=10$,$l(3,4)=10$ 
and $h=1$. 
Suppose that $f_1$ and $f_2$ have been scheduled, the new flight $f_3 \in F$, to be scheduled, is defined as:$S_3=\{1,2,3,4\}$, 
$L_3=\{(1,2),(2,3),(3,4)\}$, 
$\mathit{start}(f_3)=1$, 
$\mathit{end}(f_3)=4$, $e(f_3)=0$, $l(f_3)=21$ and 
$s(f_3)=2$.

A solution to the SD problem $(N,F)$ is represented by the pair $(R,A)$, where: (i) $R$ is a function that assigns to each flight a specific route in the network, and (ii) $A$ is an assignment of arrival times to each flight at each node along their path, such that flights are pairwise deconflicted.
A route is a sequence of nodes, pairwise connected by lanes. We write $v \in r$ and $(v,v') \in r$ to denote that node $v$ or lane $(v,v')$ are contained in the route $r=(v_1,\dots,v_n)$ that is, whenever $v=v_i$ for some $1\leq i \leq n$ or this additionally $v'=v_{i+1}$, respectively.
A route $R(f) = (v_1,\dots,v_n)$ for $f=(S,L,\mathit{start},\mathit{end},e,l,s) \in F$ has to satisfy: 
\textbf{(1)} $v_i \in S\ \forall\ i, 1\leq i \leq n$ 
\textbf{(2)} $(v_j,v_{j+1}) \in L\ \forall \ j, 1 \leq j \leq n-1$ and 
\textbf{(3)} $\mathit{start}(f) = v_1 \land \mathit{end}(f)=v_n$.
Conditions 1 and 2 
enforce routes to be connected and feasible for the flight in question and Condition 3 
ensures that each route is between a possible start and end node.
An assignment $A$ is a function $F \times V \rightarrow \mathbb{N}$, where $A(f,v)$ is undefined whenever $v \notin R(f)$. The function $A$ assigns the arrival time of a flight $f$ to a node $v$. Given a route function $R$ and $h\in\mathbb{N}$, an assignment $A$ has to satisfy the following conditions:
\textbf{(4)} $A(f,\mathit{start(f)}) \geq e(f)$ and
\textbf{(5)} $A(f,\mathit{start(f)}) \leq l(f)$.
Conditions 4 and 5
ensure that a flight starts its trip at the required departure time interval.
Next, for all $f_i, f_j \in F$ such that $\mathit{start(f_i)} = \mathit{start(f_j)}$:
\textbf{(6)} $A(f_i,\mathit{start(f_i)}) \neq A(f_j,\mathit{start(f_j)})$
\textbf{(7)} $|A(f_i,\mathit{start(f_i)}) - A(f_j,\mathit{start(f_j)})| \geq h$.
Condition 6 
ensures that flights departure times are pairwise different when flights share the same starting node.
Finally, Condition 7
ensures a safety distance between two flights that begin their trip from the same node.
For all $f=(S,L,\mathit{start},\mathit{end},e,l,s) \in F$, $R(f)=(v_1,\dots,v_n)$ such that $1\leq k \leq n$ and $h\in\mathbb{N}$
\textbf{(8)} $|A(f_i,v) - A(f_j,v)| \geq h$
\textbf{(9)} $(A(f_i,v_k) \leq A(f_j,v_k)) \land (A(f_i,v_{k+1}) \leq A(f_j,v_{k+1}))$.
Condition 8
guaranties safe distance between flights at the same node and 
Condition 9 
resolves the conflict between two flights that share the same lane.
The solution of the previous example, shown in Figure~\ref{fig:stdl-diagram}, is
$P(f_3)=(1,2,3,4)$ and $A(f_3,1)=0,A(f_3,2)=5,A(f_3,3)=10,A(f_3,4)=15$.
Strategic deconfliction involves pre-flight planning to prevent aircraft conflicts while optimizing efficiency, safety, and operational performance.
To determine the quality of a solution, we have focused our attention on minimizing total delay with respect to the required earliest launch time.
We chose this metric because it is the most widely used in the literature. Nothing prevents using a different metric or more than one to determine the best plan by assigning an evaluation priority.
The quality of a solution $(P,A)$ is determined using the following condition: $\mathit{min}\sum_{f_i \in F} A(f_i,\mathit{start(f_i)}) - e(f_i)$.
\section{Solving Real-world SD Problems with \METHOD}\label{sect:asp solution}
In this section, we present, \METHOD, our ASP-based approach to solving a couple of variants of the SD problem. 
We first show how to represent flight data, followed by the actual encoding of the problem.
\subsection{Data encoding}\label{subsect:batch fact format}
For a given SD problem $(N,F)$, the airspace network $N=(V,E,\mathit{Iv},\mathit{Ev},l,h)$ is represented by the facts
\begin{lstlisting}[language={asp}, caption={}, numbers={none},basicstyle=\ttfamily\scriptsize]
node(v). edge(v,v'). startn(iv). endn(ev). length((v,v'),l). headway(h).
\end{lstlisting}
for each $\verb|v|\in V$, $\verb|v,v'|\in E$, $\verb|iv| \in \mathit{Iv}$, $\verb|ev|\in\mathit{Ev}$, $\verb|l|\in \mathbb{N}$ and $\verb|h| \in \mathbb{N}$.

The set of flights $F$ where each $f_i=(S_i,L_i,\mathit{start_i},\mathit{end_i},e_i,l_i,s_i) \in F$ is defined by
\begin{lstlisting}[language={asp}, caption={}, numbers={none},basicstyle=\ttfamily\scriptsize, mathescape=true]
flight($f_i$). lane($f_i$,v,v'). start($f_i$,iv). end($f_i$,ev). speed($f_i$,s). requested($f_i$,e,l).
\end{lstlisting}
with \verb|flight(|$f_i$\verb|)| the flight identification and for each $\verb|v|\in S_i$, $\verb|v,v'| \in L_i$, $\mathit{start}(f_i)=\verb|iv|$, $\mathit{end}(f_i)=\verb|ev|$, $e(f_i)=\verb|e|$, $l(f_i)=\verb|l|$ and $s(f)=\verb|s|$.
For example, the following facts encode the network in Figure~\ref{fig:lane-layout} plus the headway
\begin{lstlisting}[language={asp}, caption={}, label={lst:asp network}, numbers={none},basicstyle=\ttfamily\scriptsize]
node(1..4). edge(1,2). edge(2,3). edge(3,4) startn(1). endn(4). length((1,2),10). 
length((2,3),10). length((3,4),10). headway(1).
\end{lstlisting}
while the set $F=\{f_1,f_2,f_3\}$ of flights to be scheduled, with the respective requested launch time interval, is given by
\begin{lstlisting}[language={asp}, caption={}, label={lst:asp flights batch}, numbers={none},basicstyle=\ttfamily\scriptsize]
flight(f1). speed(f1,2). lane(f1,1,2). lane(f1,2,3). lane(f1,3,4). start(f1,1). 
end(f1,4). requested(f1,1,5).
flight(f2). speed(f2,3). lane(f2,1,2). lane(f2,2,3). lane(f2,3,4). start(f2,1). 
end(f2,4). requested(f2,2,4). 
flight(f3). speed(f3,2). lane(f3,1,2). lane(f3,2,3). lane(f3,3,4). start(f3,1). 
end(f3,4). requested(f3,3,4).
\end{lstlisting}

\subsection{Problem encoding}\label{subsect:batch encoding}
In the following, we describe the general problem encoding of Listing~\ref{lst:asp batch sd const speed}.
We encode one feasible plan as an answer set.
Line 1 defines the headway with the predicate \verb|headway(h)| where \verb|h| is a constant taken as input.
In Line 2, for each flight \verb|F| is assigned a starting time point \verb|stpoint(F,X,T)| from its requested launch time interval \verb|requested(F,E,L)| at the starting node \verb|X| with \verb|start(F,X)|. This rule ensures that Conditions 4 and 5 are met.
Line 3 is an integrity constraint and eliminate the answer sets where 1) at least two flights share the same starting time point for the same starting node, and 2) there is no safe distance between flights at the same starting point. This constraint implements Conditions 6 and 7 in the opposite way, or we look for plans that do not meet the constraint.
Rules at Lines 5--7 compute the estimated time of arrival. The predicate \verb|eta(F,Y,T)| (Line 5) is true if there is a flight \verb|F| that starts its trip at time \verb|Ti| at node \verb|X|, \verb|F| travel through the lane \verb|(X,Y)| with speed \verb|S|. The arrival time is calculated taking into account the lane length \verb|length((X,Y),D)| applying the formula $T=Ti+(D/S)$. Line 7 is applied to all other nodes knowing the arrival time at the previous one.
Line 9 guarantees a safe distance between flights at the same node and Lines 10--13 resolve the conflict between two flights that share the same lane, thus implementing Condition 8 and 9 respectively.

\begin{lstlisting}[language={asp}, xleftmargin=3.5ex, caption={ASP-based encoding of the SD problem with constant speed.}, label={lst:asp batch sd const speed}, numbers={left}, basicstyle=\ttfamily\scriptsize]
headway(h).
1{stpoint(F,X,T) : T=E..L}1 :- flight(F), requested(F,E,L), start(F,X).
:- headway(H), stpoint(F1,X,T1), stpoint(F2,X,T2), F1!=F2, |T1-T2|<H.

eta(F,Y,T) :- stpoint(F,X,Ti), speed(F,S), lane(F,X,Y), length((X,Y),D), 
              T=(Ti+(D/S)).
eta(F,Y,T) :- eta(F,X,Ti), speed(F,S), lane(F,X,Y), length((X,Y),D), T=(Ti+(D/S)).

:- headway(H), eta(F1,X,T1), eta(F2,X,T2), F1!=F2, |T1-T2|<H.
:- eta(F1,X,Tx1), eta(F1,Y,Ty1), eta(F2,X,Tx2), eta(F2,Y,Ty2), lane(F1,X,Y), 
   lane(F2,X,Y), F1!=F2, Tx1<Tx2, Ty2<Ty1.
:- stpoint(F1,X,Tx1), stpoint(F2,X,Tx2), eta(F1,Y,Ty1), eta(F2,Y,Ty2), 
   lane(F1,X,Y), lane(F2,X,Y), F1!=F2, Tx1<Tx2, Ty2<Ty1.
\end{lstlisting}

Encoding of Listing~\ref{lst:asp batch sd const speed} assumes constant speed of all flights through all lanes.
For example, to avoid potential collisions with other aircraft or obstacles, speed adjustments are made to maintain safe distances or to avoid severe turbulence, speed is adjusted to reduce the impact of weather disturbances.
To manage this kind of situation it is necessary to make a simple change for the predicate \verb|speed(f,s)| into \verb|speed(f,s,(x,y))|, where \verb|f| flies at speed \verb|s| through the lane \verb|(x,y)|.
At this point, Listing~\ref{lst:asp batch sd const speed} subsumes a small variation for the rules that are necessary for estimated time of arrival.
Listing~\ref{lst:asp batch sd var speed} shows what is necessary to manage variable speed. In particular, the change is made only for 
for the predicate \verb|speed(F,S,(X,Y))|.

\begin{lstlisting}[language={asp}, xleftmargin=3.5ex, caption={ASP-based encoding of the SD problem with variable speed.}, label={lst:asp batch sd var speed}, numbers={left}, basicstyle=\ttfamily\scriptsize, firstnumber={5}]
eta(F,Y,T) :- stpoint(F,X,Ti), speed(F,S,(X,Y)), lane(F,X,Y), length((X,Y),D), 
              T=(Ti+(D/S)).
eta(F,Y,T) :- eta(F,X,Ti), speed(F,S,(X,Y)), lane(F,X,Y), length((X,Y),D), 
              T=(Ti+(D/S)).
\end{lstlisting}

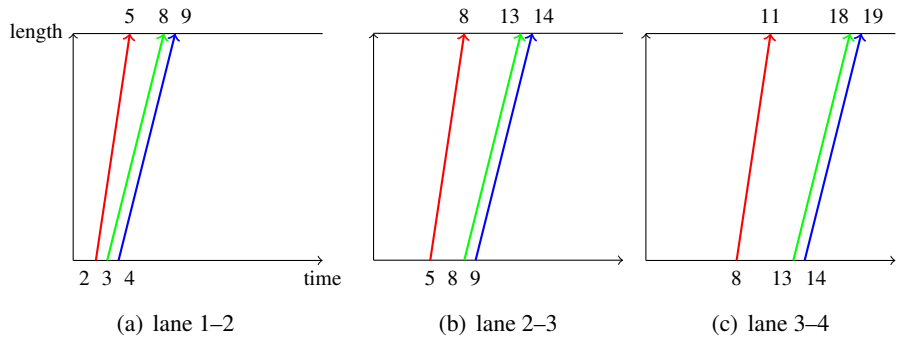
\begin{figure}[ht]
    \centering
    \subfigure[lane 1--2]{
    \begin{tikzpicture}[scale=0.30]
    \draw[->] (0,0) -- (11,0) node[right, below] {\scriptsize time};
    \draw[->] (0,0) -- (0,10) node[left] {\scriptsize length};
    \draw[] (0,10) -- (11,10);

    \draw (2.5,0) node[right, below, black] {\scriptsize 4};
    \draw[blue, thick, ->] (2,0) -- (4.5,10);
    \draw (5,10) node[right, above, black] {\scriptsize 9};
    
    \draw (0.5,0) node[right, below, black] {\scriptsize 2};
    \draw[red, thick, ->] (1,0) -- (2.5,10);
    \draw (2.5,10) node[right, above, black] {\scriptsize 5};
    
    \draw (1.5,0) node[right, below, black] {\scriptsize 3};
    \draw[green, thick, ->] (1.5,0) -- (4,10); 
    \draw(4,10) node[right, above, black] {\scriptsize 8};
    \end{tikzpicture}
    }
    \subfigure[lane 2--3]{
    \begin{tikzpicture}[scale=0.30]
    \draw[->] (0,0) -- (11,0);
    \draw[->] (0,0) -- (0,10);
    \draw[] (0,10) -- (11,10);

    \draw (4.5,0) node[right, below, black] {\scriptsize 9};
    \draw[blue, thick, ->] (4.5,0) -- (7,10);
    \draw (7.5,10) node[right, above, black] {\scriptsize 14};
    
    \draw (2.5,0) node[right, below, black] {\scriptsize 5};
    \draw[red, thick, ->] (2.5,0) -- (4,10);
    \draw (4,10) node[right, above, black] {\scriptsize 8};
    
    \draw (3.5,0) node[right, below, black] {\scriptsize 8};
    \draw[green, thick, ->] (4,0) -- (6.5,10); 
    \draw(6,10) node[right, above, black] {\scriptsize 13};
    \end{tikzpicture}
    }
    \subfigure[lane 3--4]{
    \begin{tikzpicture}[scale=0.30]
    \draw[->] (0,0) -- (11,0);
    \draw[->] (0,0) -- (0,10);
    \draw[] (0,10) -- (11,10);

    \draw (7.5,0) node[right, below, black] {\scriptsize 14};
    \draw[blue, thick, ->] (7,0) -- (9.5,10);
    \draw (10,10) node[right, above, black] {\scriptsize 19};
    
    \draw (4,0) node[right, below, black] {\scriptsize 8};
    \draw[red, thick, ->] (4,0) -- (5.5,10);
    \draw (5.5,10) node[right, above, black] {\scriptsize 11};
    
    \draw (6,0) node[right, below, black] {\scriptsize 13};
    \draw[green, thick, ->] (6.5,0) -- (9,10); 
    \draw(8.5,10) node[right, above, black] {\scriptsize 18};
    \end{tikzpicture}
    }
    \caption{STDL representation of the solution for flights \texttt{f1} (blue), \texttt{f2} (red) and \texttt{f3} (green).}
    \label{fig:solution}
\end{figure}

Listing~\ref{lst:asp batch min delay} explains the delay optimization process.
The rule at Line~1 computes the delay \verb|delay(F,D)| of a flight \verb|F| as the absolute difference between the actual launch time \verb|T| and the earliest requested launch time \verb|E|.
Line 2 is a sum of all the delays in each flight.
Finally, Line 3 gives the best scheduled plan based on the minimal delay of the entire set of flights.
As an example, for the problem instance in Section~\ref{subsect:batch fact format}, the best scheduled plan is graphically depicted in Figure~\ref{fig:solution}.

\begin{lstlisting}[language={asp}, xleftmargin=3.5ex, caption={Delay optimization.}, label={lst:asp batch min delay}, numbers={left}, basicstyle=\ttfamily\scriptsize]
delay(F,D) :- flight(F), stpoint(F,X,T), requested(F,E,L), D=T-E.
sum_delay(Sd) :- Sd = #sum{D : flight(F),delay(F,D)}.
#minimize {Sd : sum_delay(Sd)}.
\end{lstlisting}

\section{Evaluation}\label{sect:sd-evaluation}
This section presents the experimental results obtained for the three presented use cases under three UAM and synthetic square grid network topologies.
The SD problem instances are solved with the \METHOD encodings 
and also with the CP approach for comparison purposes.
We first present the experimental design, that is, the test instances and the experimental setup, including the software used and the parameters chosen.
\vspace{-5mm}
\subsection{Test instances}\label{subsect:testing instances}
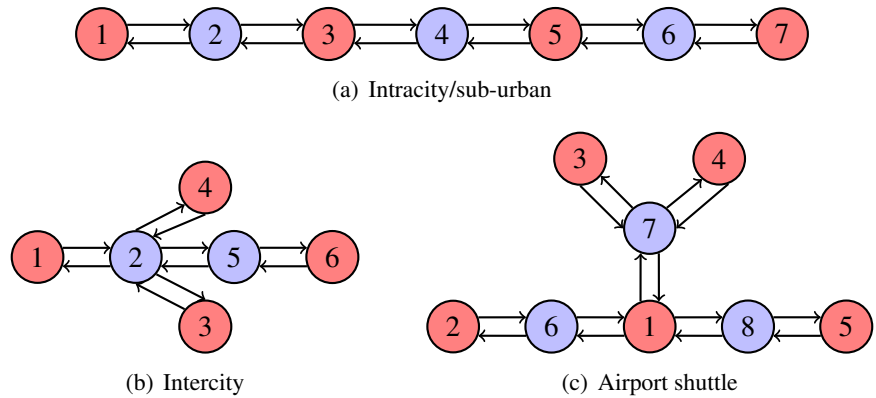
\begin{figure*}[ht]
    \centering
    \subfigure[Intracity/sub-urban]{
    \begin{tikzpicture}[scale=0.5, node distance={15mm}, thick, main/.style = {draw, circle}] 
    \node[main] (1) [fill=red!50] {1}; 
    \node[main] (2) [right of=1, fill=blue!25] {2}; 
    \node[main] (3) [right of=2, fill=red!50] {3}; 
    \node[main] (4) [right of=3, fill=blue!25] {4};
    \node[main] (5) [right of=4, fill=red!50] {5};
    \node[main] (6) [right of=5, fill=blue!25] {6};
    \node[main] (7) [right of=6, fill=red!50] {7};
    
    \draw[->] (1.20) -- (2.160);
    \draw[->] (2.200) -- (1.340);
    
    \draw[->] (2.20) -- (3.160);
    \draw[->] (3.200) -- (2.340);

    \draw[->] (3.20) -- (4.160);
    \draw[->] (4.200) -- (3.340);

    \draw[->] (4.20) -- (5.160);
    \draw[->] (5.200) -- (4.340);

    \draw[->] (5.20) -- (6.160);
    \draw[->] (6.200) -- (5.340);

    \draw[->] (6.20) -- (7.160);
    \draw[->] (7.200) -- (6.340);
    
    \end{tikzpicture}
    }
    \\
    \subfigure[Intercity]{
    \begin{tikzpicture}[scale=0.5, node distance={13mm}, thick, main/.style = {draw, circle}] 
    \node[main] (1) [fill=red!50] {1}; 
    \node[main] (2) [right of=1, fill=blue!25] {2}; 
    \node[main] (3) [below right of=2, fill=red!50] {3}; 
    \node[main] (4) [above right of=2, fill=red!50] {4};
    \node[main] (5) [right of=2, fill=blue!25] {5};
    \node[main] (6) [right of=5, fill=red!50] {6};
    
    \draw[->] (1.20) -- (2.160);
    \draw[->] (2.200) -- (1.340);
    
    \draw[->] (2.20) -- (5.160);
    \draw[->] (5.200) -- (2.340);

    \draw[->] (2.320) -- (3.90);
    \draw[->] (3.140) -- (2.270);
    
    \draw[->] (2.90) -- (4.220);
    \draw[->] (4.270) -- (2.50);

    \draw[->] (5.20) -- (6.160);
    \draw[->] (6.200) -- (5.340);

    \end{tikzpicture}
    }
    \hspace{5mm}
    \subfigure[Airport shuttle]{
    \begin{tikzpicture}[scale=0.5, node distance={13mm}, thick, main/.style = {draw, circle}] 
    \node[main] (2) [fill=red!50] {2};
    \node[main] (6) [right of=2, fill=blue!25] {6};
    \node[main] (1) [right of=6, fill=red!50] {1}; 
    \node[main] (8) [right of=1, fill=blue!25] {8};
    \node[main] (5) [right of=8, fill=red!50] {5};
    \node[main] (7) [above of=1, fill=blue!25] {7};
    \node[main] (3) [above left of=7, fill=red!50] {3};
    \node[main] (4) [above right of=7, fill=red!50] {4};
    
    \draw[->] (1.20) -- (8.160);
    \draw[->] (8.200) -- (1.340);

    \draw[->] (8.20) -- (5.160);
    \draw[->] (5.200) -- (8.340);

    \draw[->] (6.20) -- (1.160);
    \draw[->] (1.200) -- (6.340);

    \draw[->] (2.20) -- (6.160);
    \draw[->] (6.200) -- (2.340);

    \draw[->] (1.110) -- (7.250);
    \draw[->] (7.290) -- (1.70);

    \draw[->] (7.130) -- (3.320);
    \draw[->] (3.270) -- (7.180);

    \draw[->] (7.50) -- (4.230);
    \draw[->] (4.290) -- (7.0);
    \end{tikzpicture}
    }
    \caption{Air network topologies.}
    \label{fig:air topologies}
\end{figure*}

We consider three UAM network topologies, illustrated in Figure~\ref{fig:air topologies}.
These represent different types of layout that might be encountered in future UAM cases \cite{asmer2021urban}.
\emph{Intracity/sub-urban} focuses on air-based transportation within a city's core, addressing urban congestion and enabling rapid point-to-point mobility. 
\emph{Intercity} focuses on air-based transportation between cities or major regional hubs, filling gaps between traditional ground transit and commercial aviation.
\emph{Airport shuttle} transports passengers between airports and urban centers, nearby cities, or transit hubs. This use case targets time-sensitive travellers seeking to bypass ground congestion and streamline airport access. 
The topologies are modelled as directed graphs, where ground nodes (vertiport or vertistop) are coloured in red, while blue nodes represent waypoints up in the air.
Figure~\ref{fig:air topologies}~(a) represents a single city with multiple stops that connects the city center with suburban areas like a ``subway in the sky''.
Figure~\ref{fig:air topologies}~(b) represents different cities that are connected, and Figure~\ref{fig:air topologies}~(c) represents an airport (node 1) that serves multiple cities.
We have also defined an ASP way to generate synthetic square grid air network topologies.
We did it for the evaluation of the scalability as the size of the network increases in the number of nodes and edges. The interested reader can find details about how the different air network topologies have been generated in the Supplementary Material\footnote{\url{https://figshare.com/s/00cf3c418a014fcc51f1}}.

\vspace{-5mm}
\subsection{Experimental setup}\label{subsect:experimental setup}
All experiments were run on a laptop computer with Ubuntu 20.04.4, AMD Ryzen 5 3500U @ 2.10 GHz and 8GB RAM. As solvers we used \verb|clingo| \cite{DBLP:journals/tplp/GebserKKS19} for ASP
and \verb|Minizinc| (Gecode solver) \cite{DBLP:conf/cp/NethercoteSBBDT07} for CP. We set default solving parameters without multi-threading mode of all solvers; only the timeout has been set to 900 seconds (15 minutes).
The evaluation focuses on efficiency (time and memory requirements), as well as effectiveness (ability to find a suitable plan to schedule flights). Specifically, we define the following research questions. 
\textbf{(RQ1)} How efficient is \METHOD by varying the number of flights, the size of the launch interval and the size of the air network topology?
\textbf{(RQ2)} How efficient is \METHOD, the ASP logical approach compared to the CP approach?
To answer the RQs, we evaluated the encodings for the three air network topologies and synthetic square grid layouts.
In the absence of public datasets used as benchmarks, we were forced to generate data for the layouts.
The Supplementary Material explains how data were generated.
The launch interval requested for each flight has been chosen to follow a uniform distribution given a set of parameters such as the time horizon and min-max launch interval size.
We assume that the flight departure occurs at the beginning of the minute.
We consider vehicle speed expressed in meters per second, time and headway in minutes, and lane length in meters.
\subsection{Varying the number of flights and the take-off interval}\label{subsect:results and analysis}
\begin{figure*}[!ht]
    \centering
    \subfigure[Intracity/sub-urban]{\includegraphics[width=0.45\linewidth]{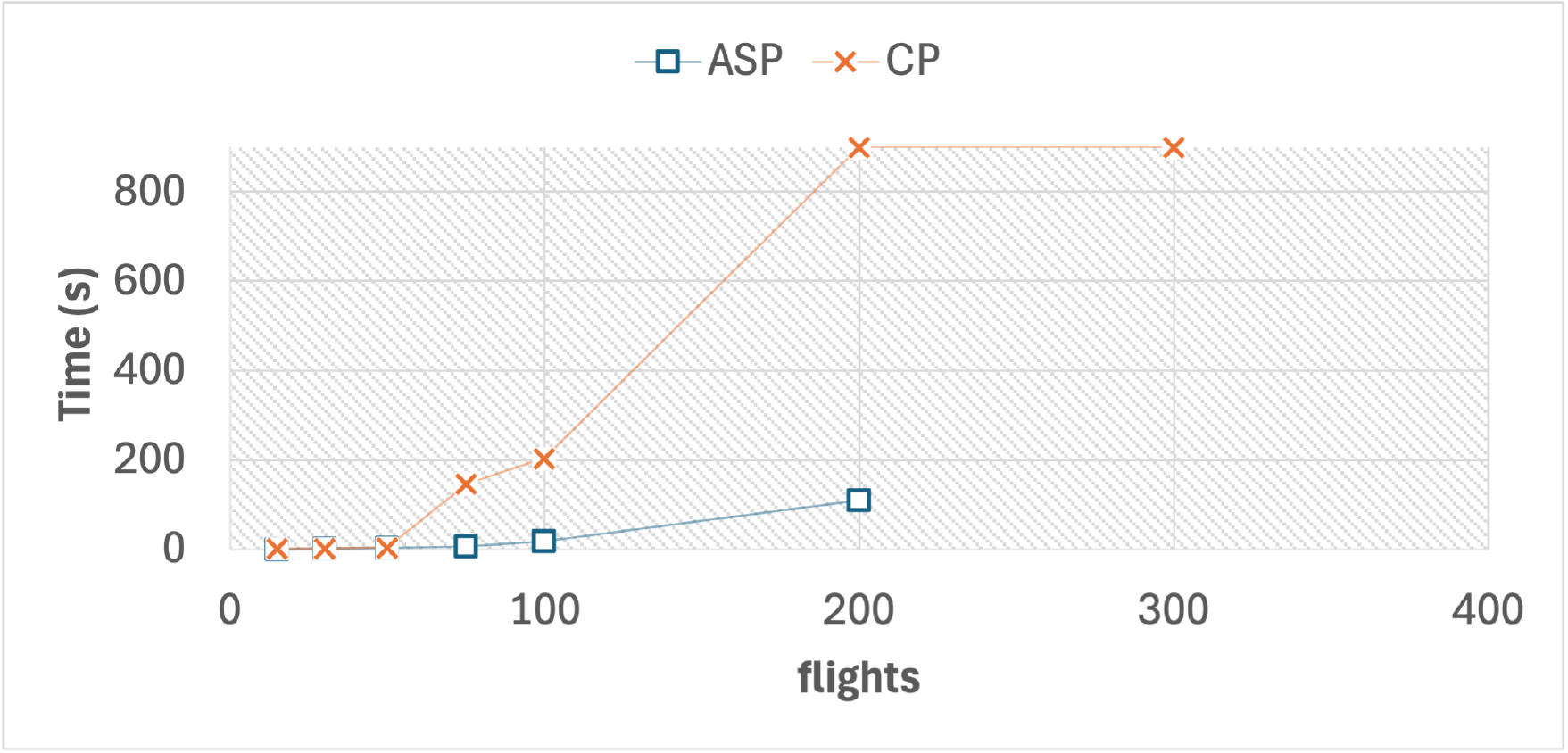}
    \includegraphics[width=0.45\linewidth]{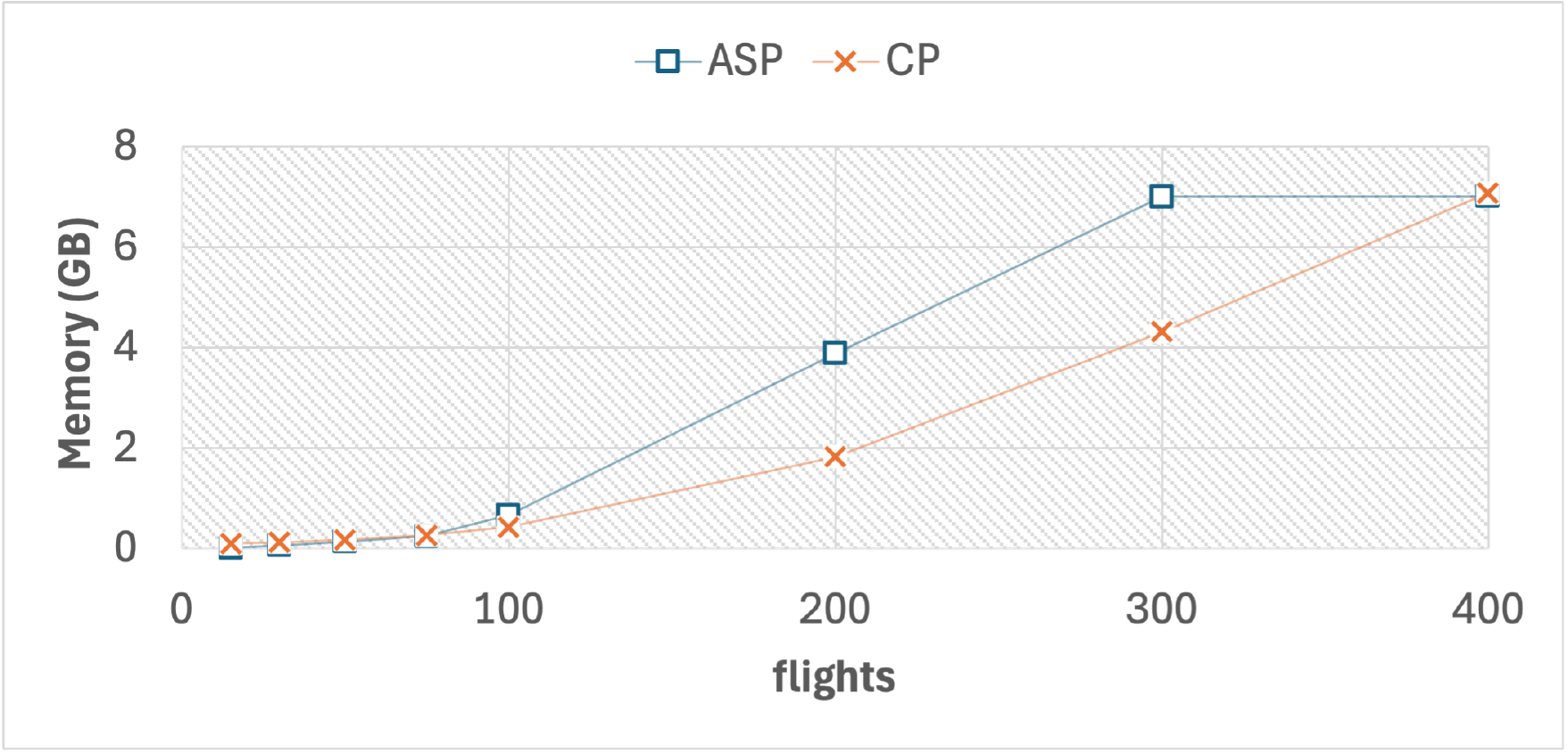}}
    \subfigure[Intercity]{\includegraphics[width=0.45\linewidth]{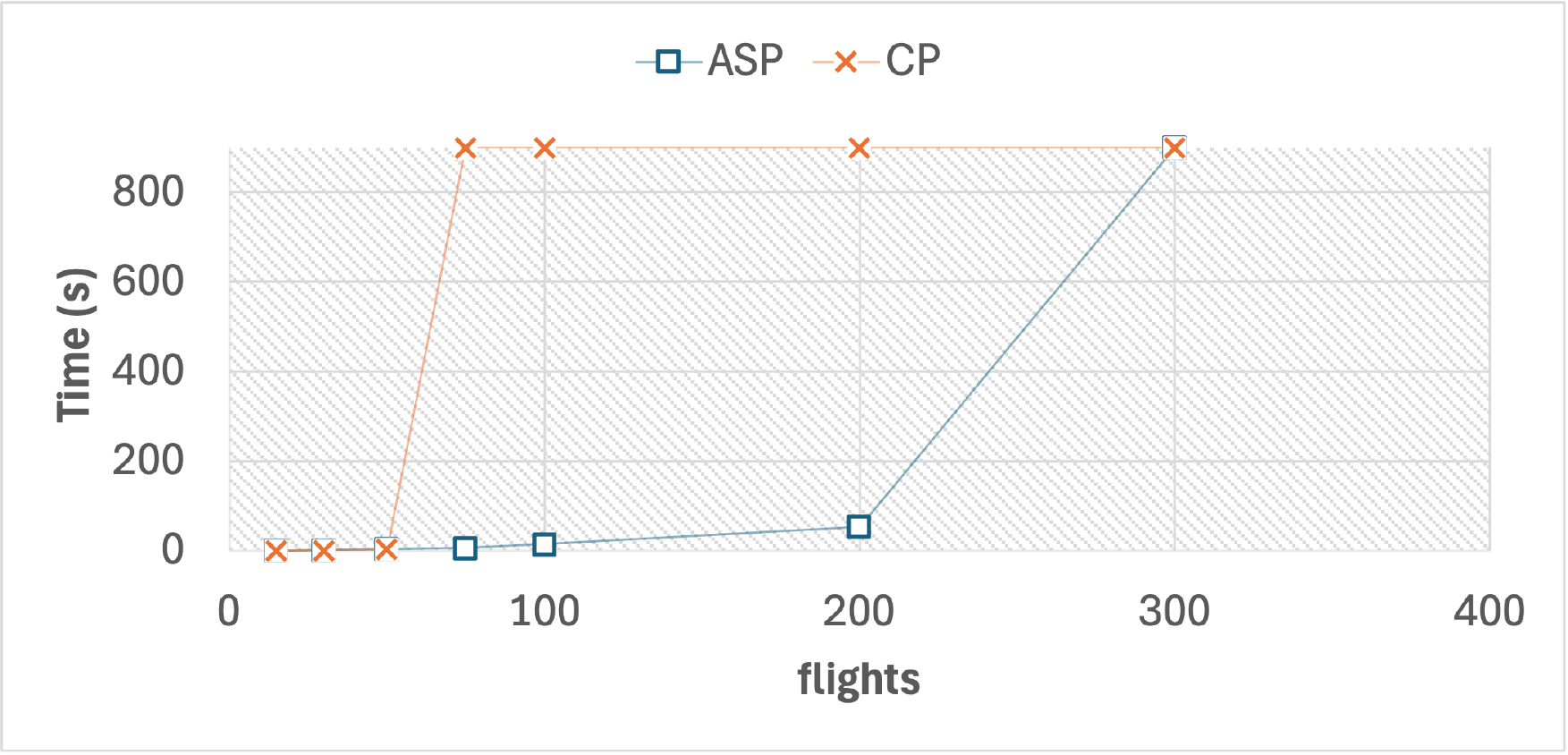}
    \includegraphics[width=0.45\linewidth]{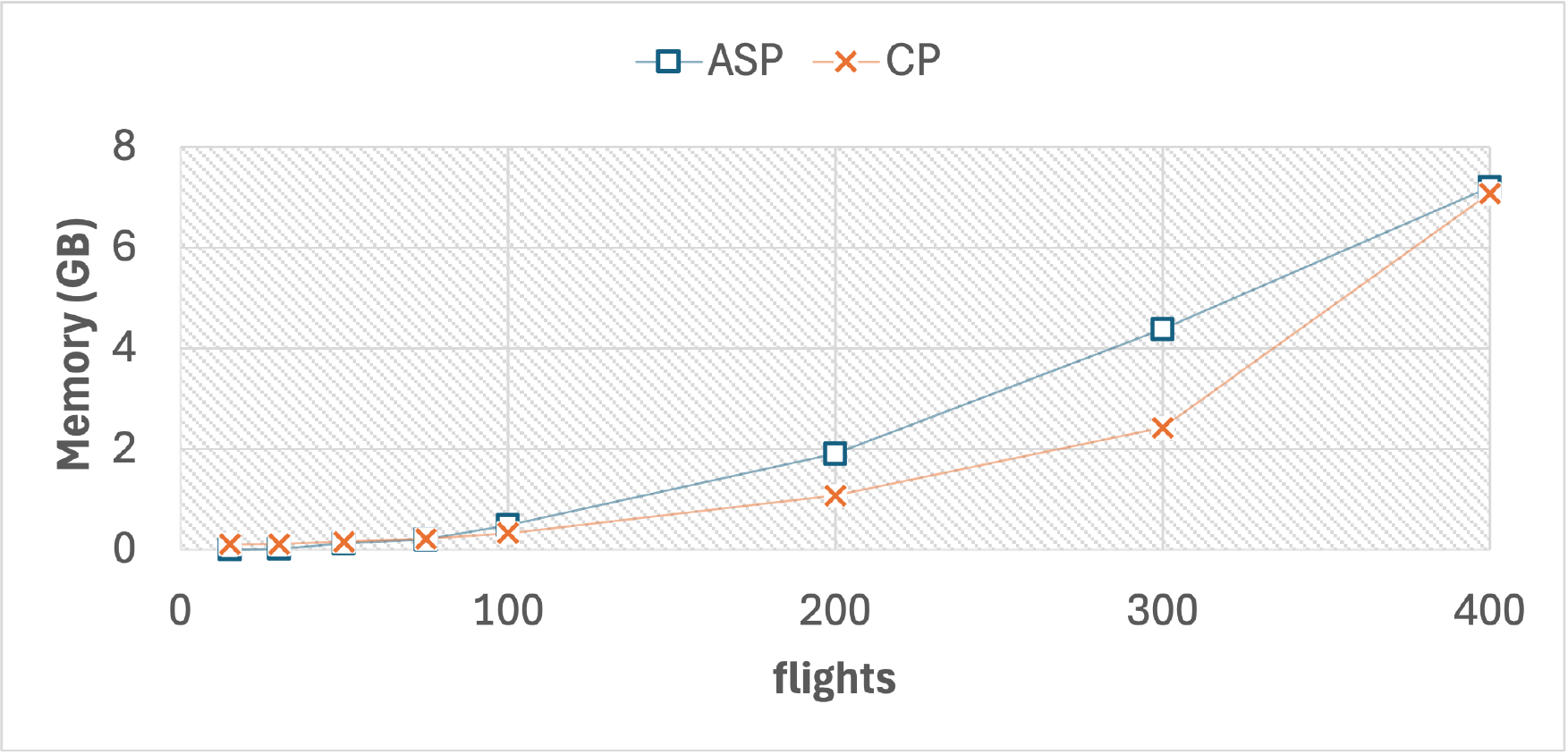}}
    \subfigure[Airport shuttle]{\includegraphics[width=0.45\linewidth]{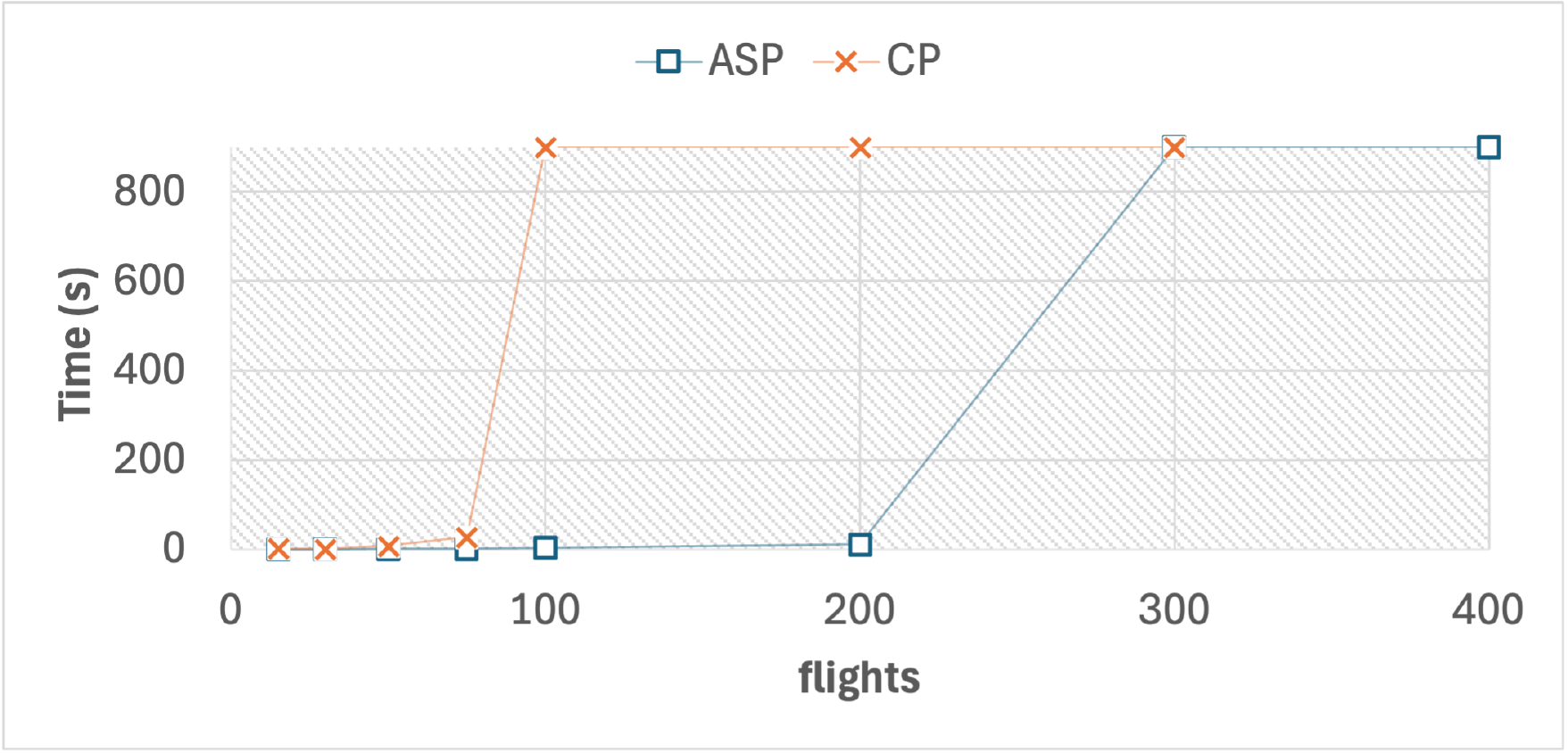}
    \includegraphics[width=0.45\linewidth]{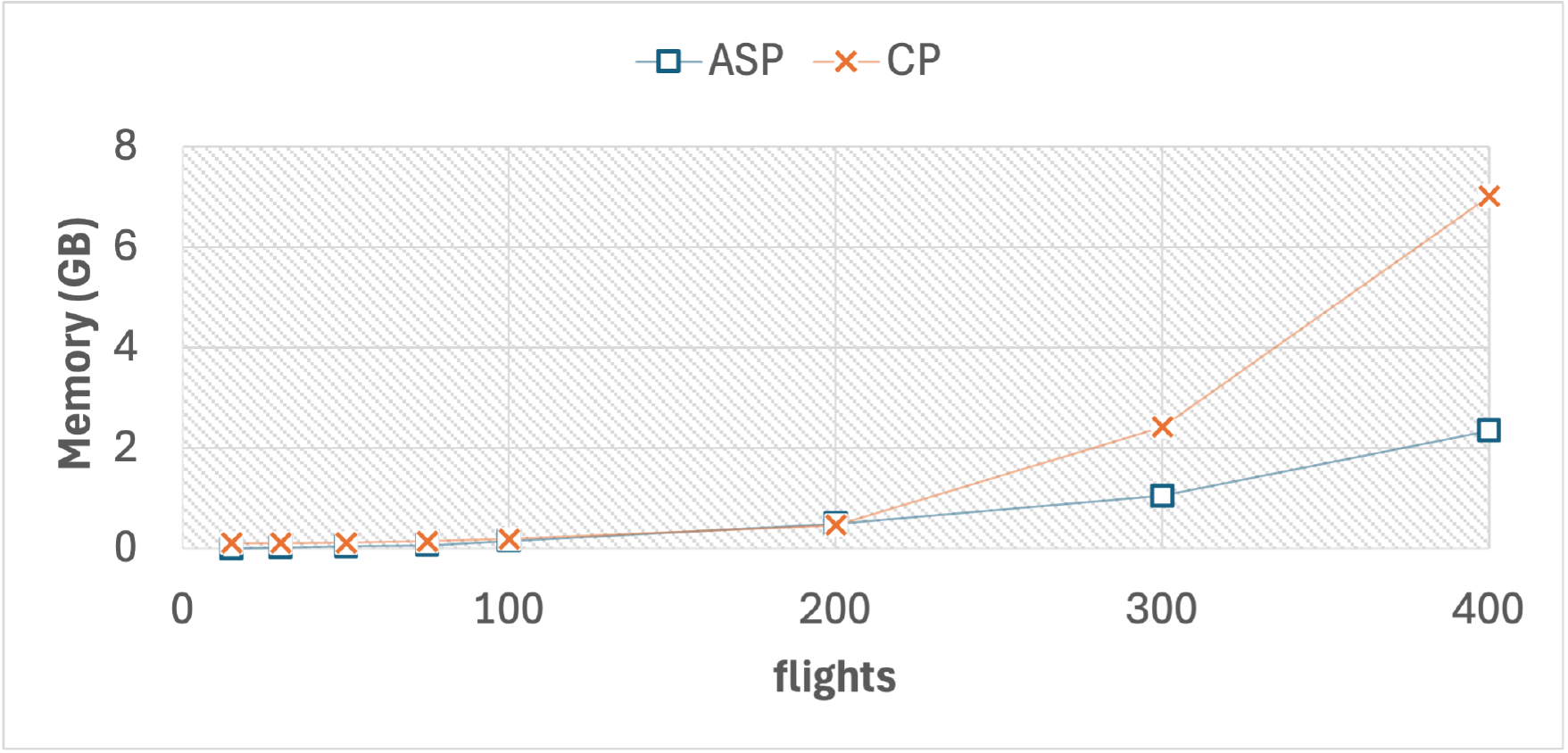}}
    \caption{Time and memory requirements varying the number of flights of the ASP approach compared with the CP approach.}
    \label{fig:varying flights results}
\end{figure*}

In the following section, 
we evaluate and discuss the proposed approach 
by varying the different parameters of the problem to understand the time and space requirements.
The evaluation of our approach is based on analysing its efficiency through scalability tests by increasing both the number of flights to be scheduled and the size of the required launch interval. 
These two dimensions are also addressed by comparing \METHOD with CP (the model is available in the Supplementary Material) in order to understand the advantages and disadvantages of the two approaches.
Figure~\ref{fig:varying flights results} shows the results obtained by varying the number of flights. 
For this evaluation, the number of flights was increased from 15 to 400 over a 6-hours time horizon. The minimum headway is 1 minute with a required launch interval size ranging between 5 and 15 minutes.
The Figure shows the time and memory requirements for the two approaches, ASP and CP, varying the number of flights in three scenarios: airport shuttle, intercity and intracity/sub-urban.
When there is no marker, it means that an out-of-memory occurred during the computation.
ASP generally demonstrates greater efficiency in terms of execution time and memory consumption up to a certain number of flights. However, it struggles with very high numbers of flights (from 300 on), often exceeding the time limit or running out of memory. 
Conversely, CP tends to require more time and less memory compared to ASP and faces significant challenges with small numbers of flights (from 75 in the intercity scenario), frequently exceeding the time limit or running out of memory earlier than ASP. 
Overall, ASP appears to be more scalable and efficient in time while CP exhibits a more consistent memory usage but struggles significantly with the execution time as the problem complexity increases.
However, we remind that the hardware used makes use of only 8 GB of memory. Therefore, with larger amounts of memory, we might not have any difficulties with the ASP approach.
Furthermore, in the configuration used by the solvers, we did not use multi-threading.

\begin{figure*}[!ht]
    \centering
    \subfigure[Intracity/sub-urban]{\includegraphics[width=0.45\linewidth]{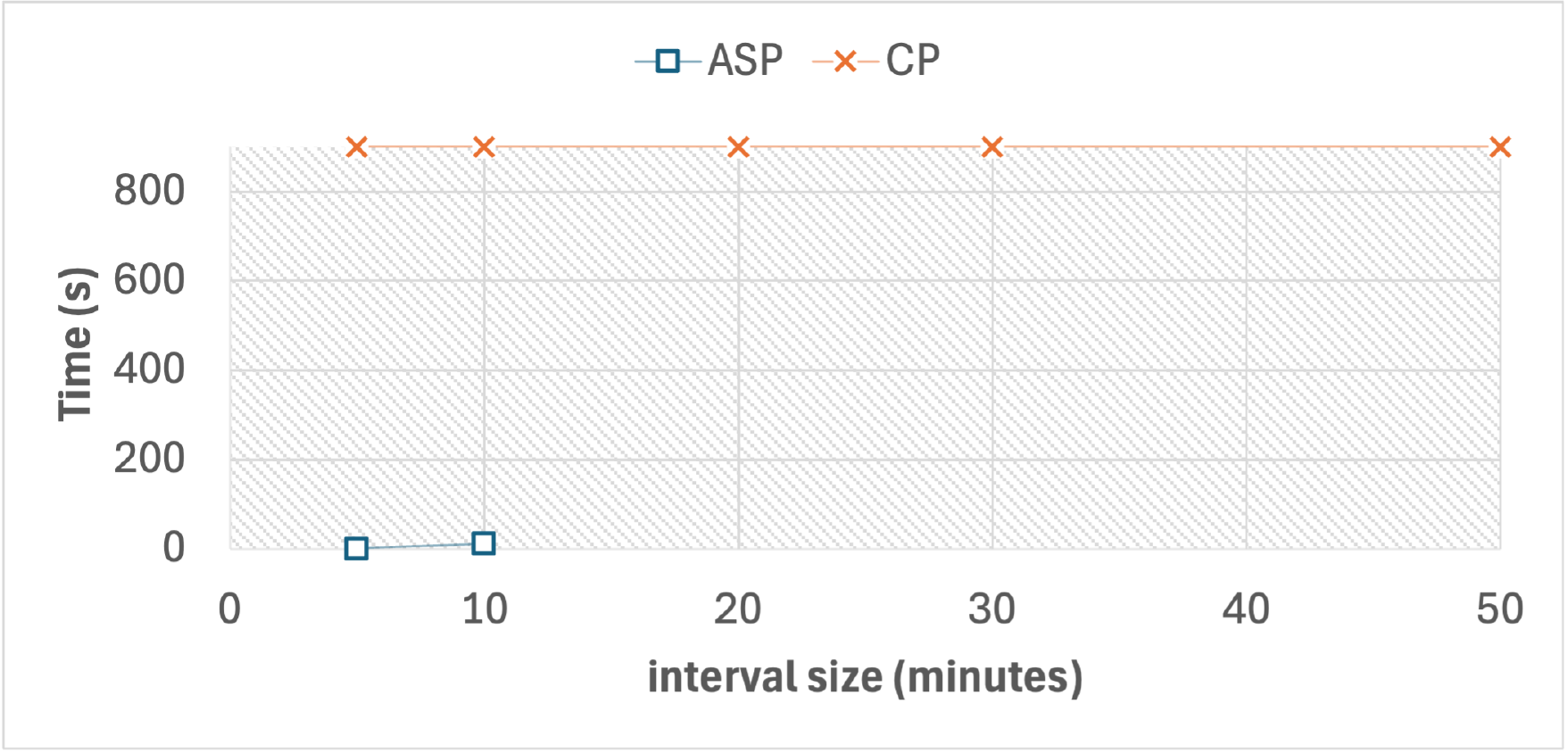}
    \includegraphics[width=0.45\linewidth]{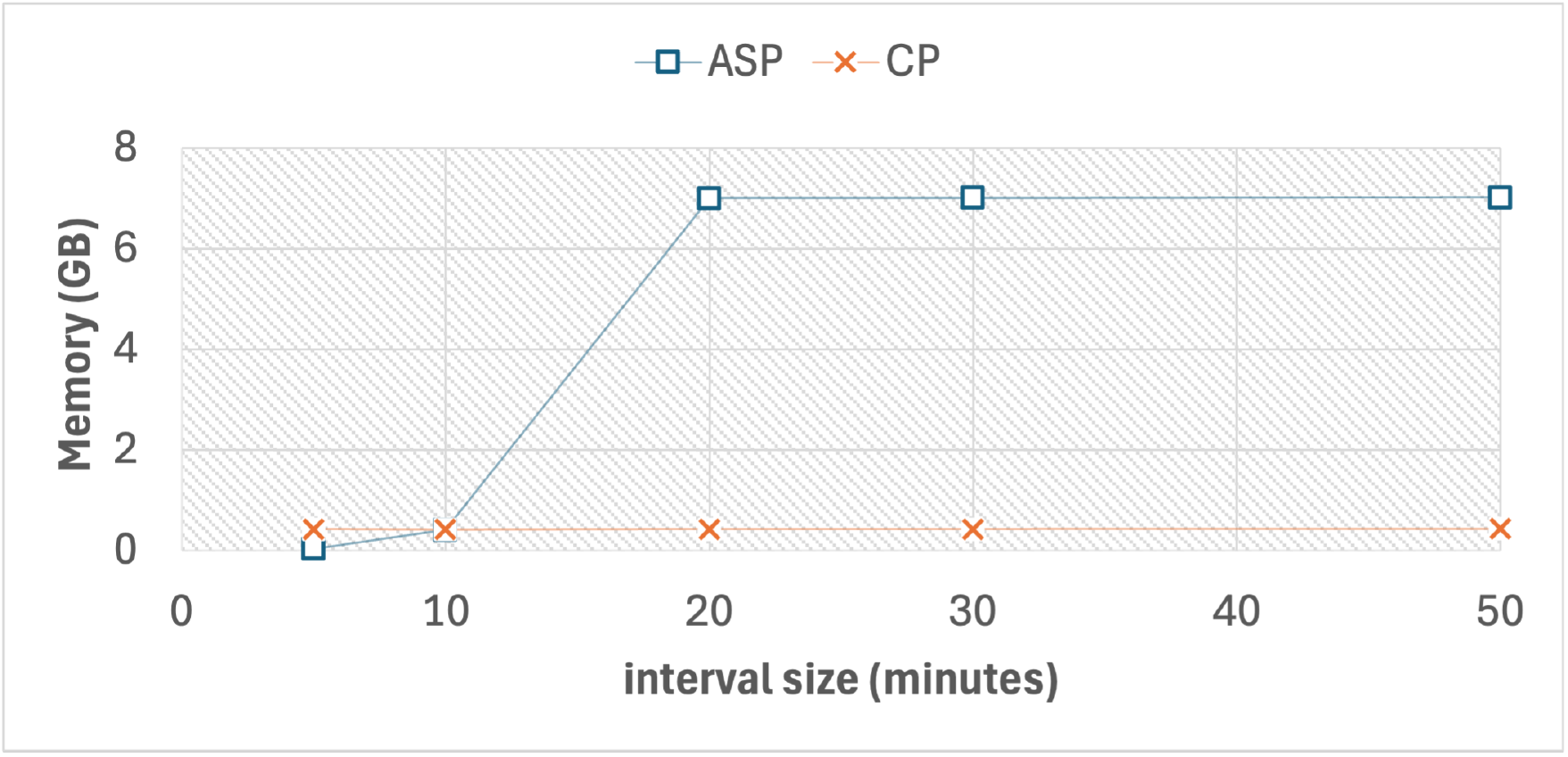}}
    \subfigure[Intercity]{\includegraphics[width=0.45\linewidth]{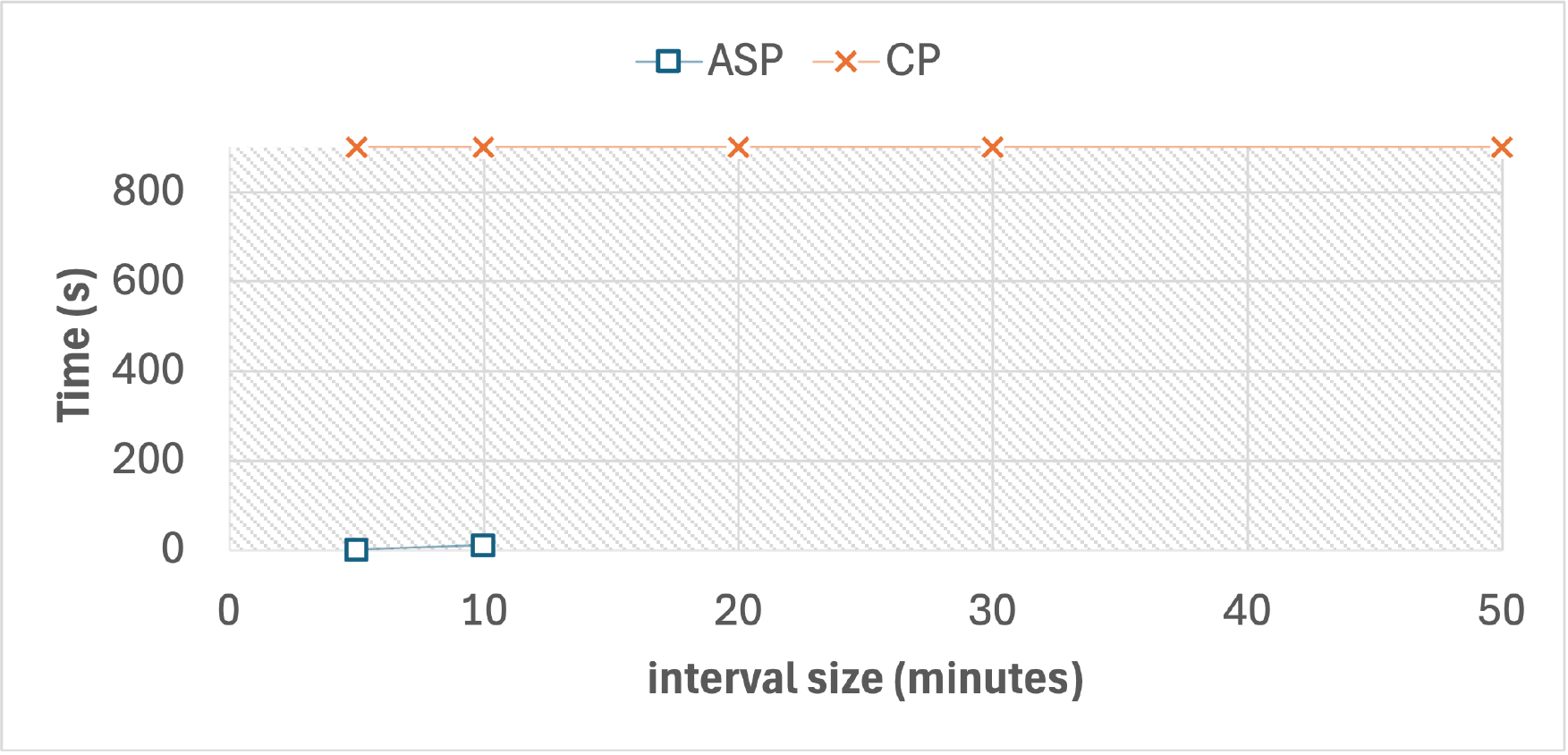}
    \includegraphics[width=0.45\linewidth]{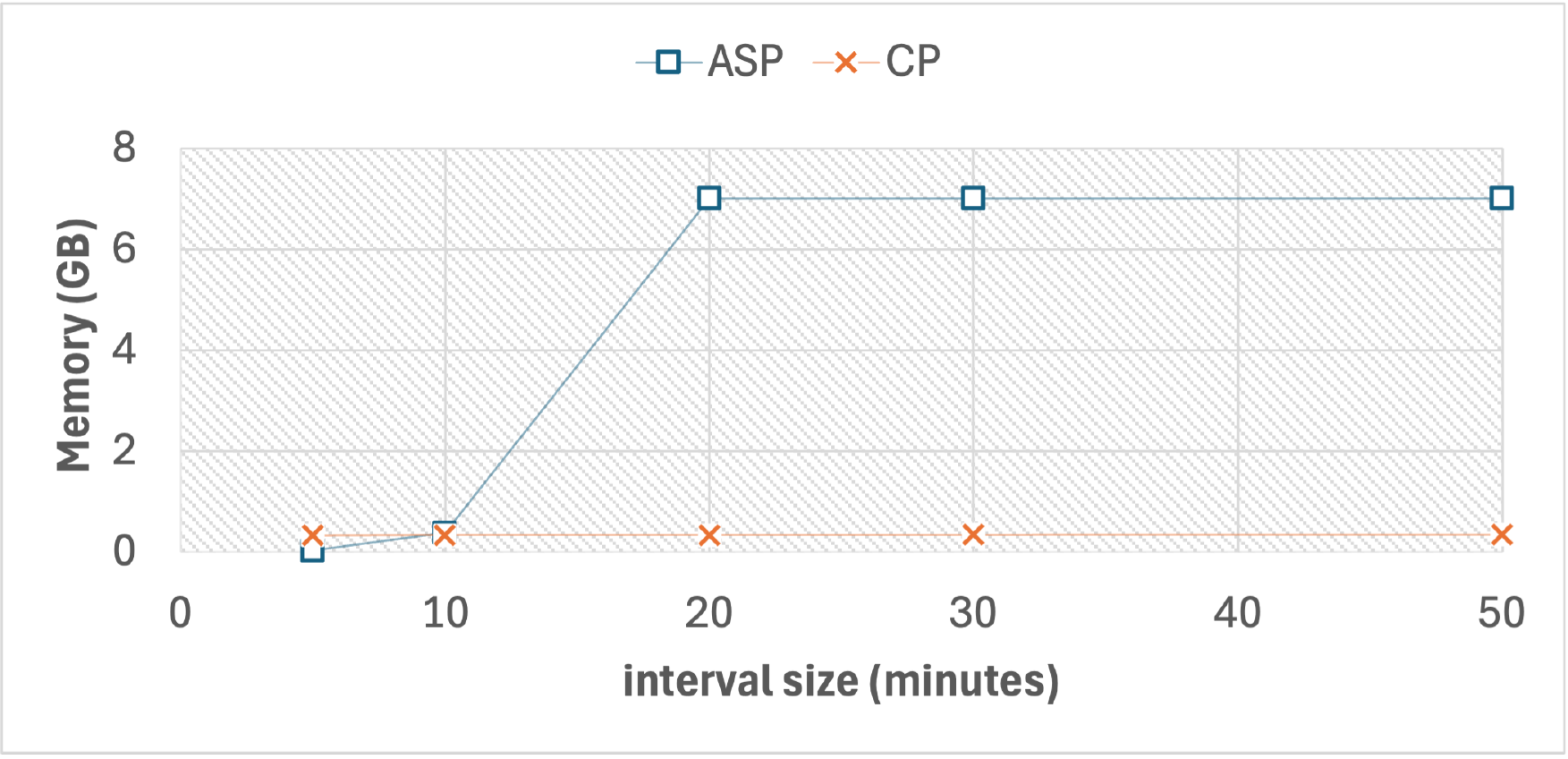}}
    \subfigure[Airport shuttle]{\includegraphics[width=0.45\linewidth]{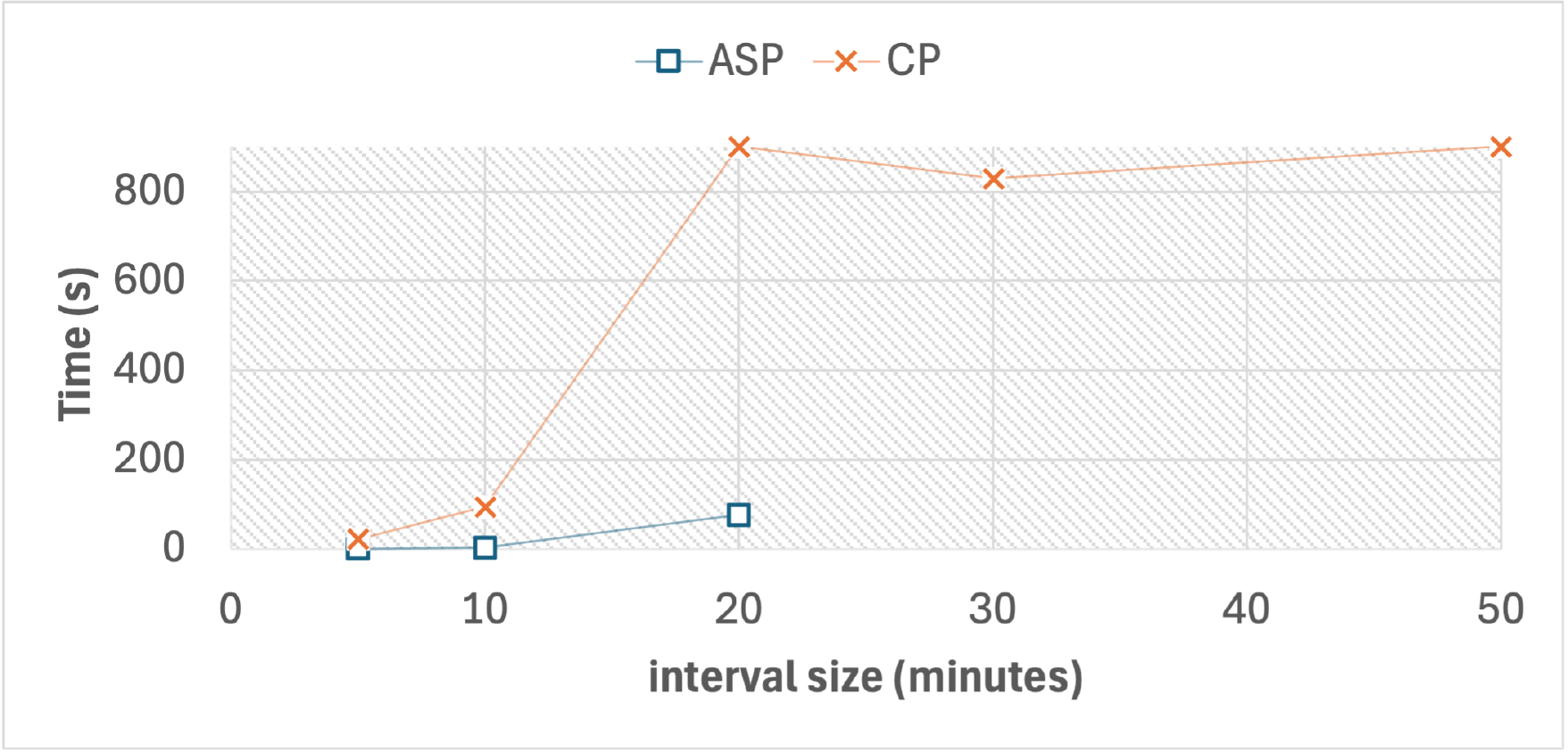}
    \includegraphics[width=0.45\linewidth]{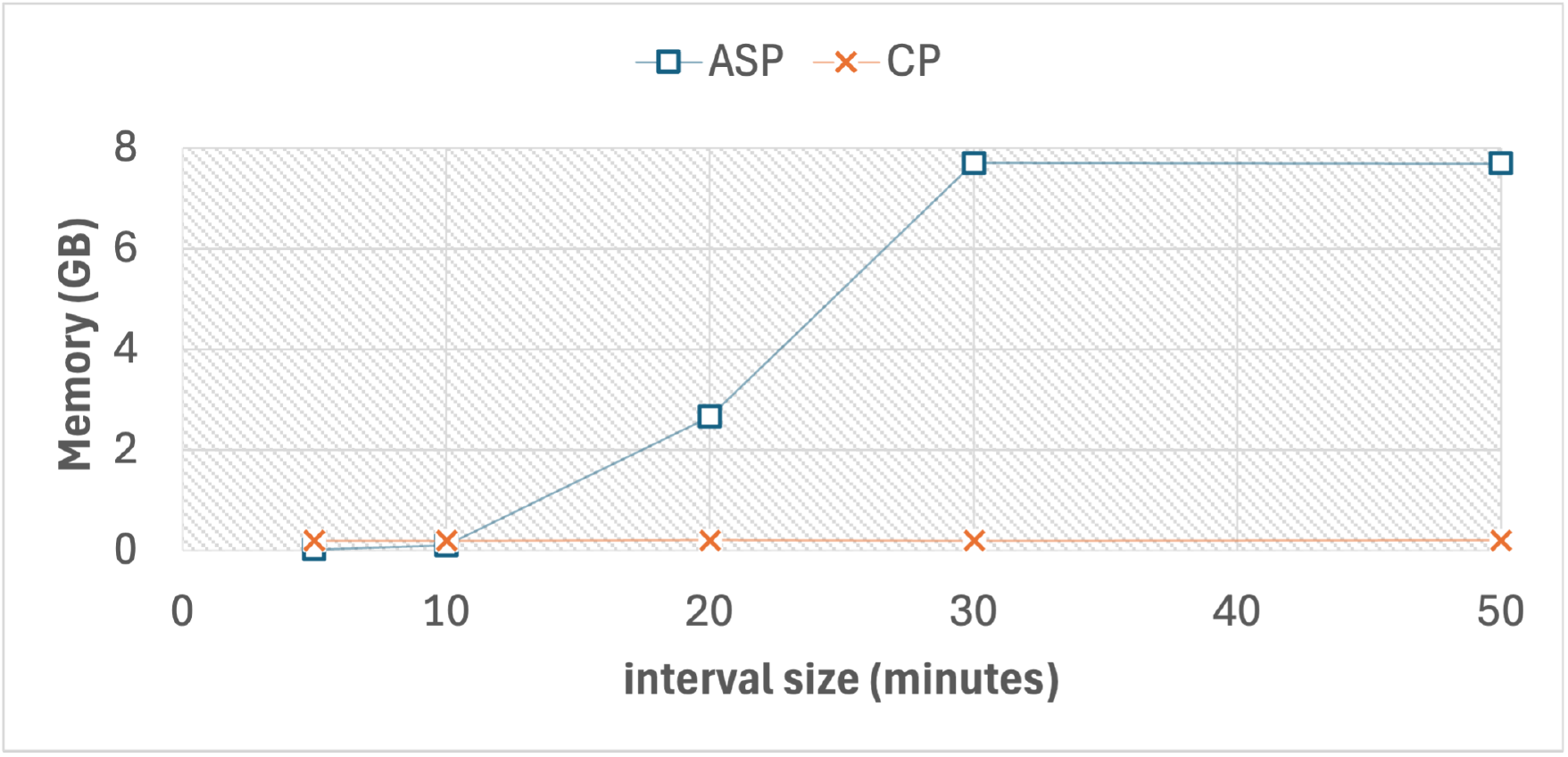}}
    \caption{Time and memory requirements varying the launch interval size of the ASP approach compared with the CP approach.}
    \label{fig:varying interval size results}
\end{figure*}

We now analyse the results shown in Figure~\ref{fig:varying interval size results}. 
We left the number of flights constant at 100, while varying the size of the required launch interval from 5 to 50 minutes.
ASP is more efficient in terms of execution time at smaller interval sizes, but struggles with larger interval sizes, often running out of memory.
CP requires significantly more time. At larger interval sizes, CP fails to find the optimum within the time limit, though it manages memory usage better than ASP.
Overall, ASP appears to be more suitable for scenarios with smaller interval sizes, while CP may be more robust in terms of memory management but less efficient in terms of execution time.

ASP is generally more efficient in terms of execution time for smaller problem sizes (both in terms of the number of flights and interval sizes). However, as the problem size increases, ASP struggles to find the optimal solution within the memory limit, especially for larger interval sizes and higher numbers of flights.
CP tends to require significantly more time to find the optimal solution compared to ASP, even for smaller problem sizes. CP consistently hits the time limit for larger interval sizes and higher numbers of flights. Memory consumption remains relatively stable across different problem sizes. CP manages memory usage better than ASP, even though it struggles with execution time.
\vspace{-5mm}
\subsection{Varying the size of the air network}\label{subsect:results-size-network}
The focus on the following section is the evaluation of \METHOD by varying the size of the air network topologies.
We used the ASP generator 
to generate square grids of sizes 3x3, 4x4, and 5x5.
The minimum headway is 1 minute with a required launch interval size of 3 minutes.
We measured time and memory requirements as in the previous section.
The results of the evaluation are showed in Table \ref{tab:time-varying-network-size}.
The experimental results demonstrate a clear correlation between the increase in problem instances (Flights) and the consumption of computational resources (time and memory).
The execution time exhibits a linear growth pattern as the number of flights increases. Specifically, in the 3×3 grid, increasing the flight load from 500 to 5,000 (a tenfold increase) results in an execution time jump from 2.26\unit{s} to 248.33\unit{s}.
As the grid expands, the execution time for the same number of flights increases. For instance, at 3,000 flights, the execution time rises from 83.35\unit{s} (3×3) to 165.16\unit{s} (5×5). This is likely due to the increased number of atoms and rules generated during the grounding phase to represent the larger spatial domain.
The memory consumption remains relatively stable and efficient for most configurations, staying well below 100 MB. However, a critical threshold is reached at the 5,000 flights-5×5 grid configuration.
The peak memory jumps dramatically to 7.10 GB. This exponential surge indicates a ``combinatorial explosion'', typical in ASP when the grounder generates a massive number of constraints that exceed the available memory.
The results confirm that the proposed ASP approach is highly viable for real-time or near-real-time strategic deconfliction in UAM environments up to a certain density. The transition from 79.72 MB to 7.10 GB at the highest tested density highlights the limit of scalability for the current model formulation, suggesting that for larger networks or higher flight volumes, a more suitable representations may be required.

\begin{table}[!ht]
    \centering
    \caption{Execution time and memory peak. Only memory peak in case of out of memory.}
    \label{tab:time-varying-network-size}
    \begin{tabular}{rrrr}
    \hline
    Flights & 3x3 & 4x4 & 5x5 \\
    \hline
    500 & 2.26 \unit{s}, 16.36 MB & 3.37\unit{s}, 17.36 MB & 4.36\unit{s}, 17.93 MB\\
    1000 & 8.85\unit{s}, 21.18 MB & 13.08\unit{s}, 20.55 MB & 17.44\unit{s}, 22.51 MB\\
    1500 & 20.02\unit{s}, 22.26 MB & 29.00\unit{s}, 30.00 MB & 40.29\unit{s}, 26.90 MB\\
    2000 & 36.07\unit{s}, 32.62 MB & 52.14\unit{s}, 31.14 MB & 71.76\unit{s}, 38.29 MB\\
    3000 & 83.35\unit{s}, 39.72 MB & 124.22\unit{s}, 49.73 MB & 165.16\unit{s}, 53.09 MB\\
    5000 & 248.33\unit{s}, 62.36 MB & 369.48\unit{s}, 79.72 MB & 7.10 GB \\
    \hline
    \end{tabular}
\end{table}

\vspace{-10mm}
\section{Conclusions}\label{sect:sd-conclusions}
To the best of our knowledge, this work introduces the first ASP-based formulation for UAM Strategic Deconfliction, named \METHOD. We define conditions to ensure separation in UAM settings and use ASP to model the air network topology, drone fleet, and the SD problem. ASP provides key advantages: \textbf{(I)} simplified and readable modelling, \textbf{(II)} support for complex combinatorial optimization, and \textbf{(III)} adaptability to specification changes without major reprogramming. Our model accommodates diverse operational UAM scenarios within realistic topologies envisioned for future services.
Our study examines the capabilities and limitations of ASP and CP in strategic deconfliction under varying conditions. It highlights key factors influencing SD complexity. ASP offers superior time efficiency for smaller problems, defined by fewer flights and narrower intervals. However, as problem size grows, ASP struggles to find optimal solutions within memory constraints, especially with larger intervals and higher flight counts. In contrast, CP requires significantly more time than ASP, even for small problems, and often hits time limits for larger cases. Nonetheless, CP maintains stable and efficient memory usage across problem sizes, outperforming ASP in this regard despite its slower execution.

While this study demonstrates the efficacy of ASP-based strategic deconfliction within a controlled environment, several avenues for future research remain to broaden these findings. To move beyond the current synthetic instances, future work will focus on validating the model against real-world datasets and diverse operational scenarios. Furthermore, to ensure the findings are not hardware-dependent, we intend to conduct an extensive cross-platform analysis involving varied computational architectures.
A critical next step involves expanding the methodological scope by integrating multiple ASP solvers such as clingo[DL] and ALASPO and comparing their performance against a wider array of external optimization stacks beyond the MiniZinc/Gecode framework. 
In addition to hardware and solver diversification, future evaluations will focus on scaling the problem dimensions by increasing the number of aircraft and expanding the complexity of airspace topologies. These experiments are designed to probe the operational limits and computational stress points of the proposed architecture, providing a definitive assessment of its scalability in high-density scenarios. Nevertheless, it should be noted that the flight volumes and environmental configurations employed in our current simulations remain strictly aligned with contemporary UAM benchmarks and existing literature. Given that the primary challenges of this aviation sector reside in constrained urban environments, the current experimental parameters accurately reflect the practical demands and deployment realities of real-world strategic deconfliction.
By diversifying the experimental parameters and software environments, we aim to establish more robust conclusions regarding the general suitability and scalability of ASP for complex strategic deconfliction tasks.

\bibliographystyle{eptcs}
\bibliography{bib/asp-scheduling,bib/asp,bib/constraint-programming,bib/deconfliction,bib/uam,bib/uav-applications}
\end{document}